\documentclass[preprint,aps]{revtex4}
\usepackage{amsmath}
\usepackage{amssymb}

\input{tcilatex}

\begin{document}

\title{Rigorous description of exchange-correlation energy of many-electron
systems}
\author{Yu-Liang Liu}
\affiliation{Department of Physics, Renmin University of China, Beijing 100872, People's
Republic of China}

\begin{abstract}
With the eigenfunctional theory, we study a general interacting electron
system, and give a rigorous expression of its ground state energy which is
composed of two parts, one part is contributed by the non-interacting
electrons, and another one is represented by the correlation functions that
are controlled by the electron correlation. Moreover, according to the
rigorous expression of the ground state energy, an effective method beyond
the local density approximation of the density functional theory is proposed.

74.72.-h.
\end{abstract}

\pacs{}
\maketitle

The electron correlation plays an important role in description of
many-electron systems. According to the electron correlation strength, the
systems are approximately divided into two categories, one is called weakly
correlated electron systems, and another one is called strongly correlated
electron system\cite{1}. The former can be approximately described by the
independent-particle (quasiparticle) schemes\cite{2,3}, while the latter has
to be represented by new schemes beyond the independent-particle description.

The Hohenberg-Kohn-Sham (HKS) density functional theory\cite{4,5} opens a
new era in description of the many-electron systems, and it and its
generalizations\cite{6} are extensively applied in physics and chemistry.
However, the rigorous expression of the exchange-correlation energy in the
HKS density functional theory is unknown, and the local density
approximation (LDA) of the Kohn-Sham scheme plays a central role in
calculating the ground state energy of the systems. In general, for the
weakly correlated systems, the result of the LDA is accurately agreement
with experiment data, but it is unreliable for the strongly correlated
systems because of the strong electron correlation.

In contrast to the Kohn-Sham scheme of the density functional theory, we
study a general quantum many-electron system with the eigenfunctional theory%
\cite{7}, and give a rigorous expression of the ground state energy,
especially the correlation energy. Moreover, according to this rigorous form
of the ground state energy, the correlation energy can be well defined, and
a more effective method beyond the LDA is proposed.

The Hamiltonian of a many-electron system can be generally written down,%
\begin{equation}
\widehat{H}=\widehat{H}_{0}+\frac{e^{2}}{2}\int d^{3}xd^{3}x^{\prime
}V(x-x^{\prime })\widehat{\rho }(x)\widehat{\rho }(x^{\prime })  \label{1}
\end{equation}%
where $\widehat{H}_{0}=\sum_{\sigma }\int dx\widehat{\psi }_{\sigma
}^{\dagger }(x)\left( \frac{\widehat{p}^{2}}{2m}+U(x)\right) \widehat{\psi }%
_{\sigma }(x)$ is the Hamiltonian of the non-interacting electrons, $%
\widehat{\psi }_{\sigma }^{\dagger }(x)$ ($\widehat{\psi }_{\sigma }(x)$)
are the creation (annihilation) operators of the electrons with spin label $%
\sigma $ at the coordinate $\mathbf{x}$, $\widehat{\rho }(x)=\sum_{\sigma }%
\widehat{\psi }_{\sigma }^{\dagger }(x)\widehat{\psi }_{\sigma }(x)$ is the
density operator of the electrons, and $U(x)$ is an external potential. The
last term represents the electron Coulomb interaction, and it induces the
electron correlation effect which makes the problem hard be treated. In
usual mean field theory\cite{8,9}, the (effective) Coulomb interaction is
treated as a perturbation parameter, while in the density functional theory,
it is incorporated into the exchange-correlation energy. However, if the
electron correlation is strong, in which case the results obtained by the
mean field theory and the LDA of the density functional theory are
unreliable, and the electron Coulomb interaction has to be accurately
treated.

With the eigenfunctional theory, the partition function of the system reads%
\cite{7},%
\begin{eqnarray}
Z &=&\int D\psi ^{\ast }D\psi D\rho D\phi e^{\frac{i}{\hbar }S}  \notag \\
S &=&\sum_{\sigma }\int dxdt\psi _{\sigma }^{\ast }(x,t)\widehat{M}(x,t)\psi
_{\sigma }(x,t)+W[\rho ,\phi ]  \label{2}
\end{eqnarray}%
where $W[\rho ,\phi ]=\int dxdt\phi (x,t)\rho (x,t)-\frac{e^{2}}{2}\int
dtd^{3}xd^{3}x^{\prime }V(x-x^{\prime })\rho (x,t)\rho (x^{\prime },t)$, $%
\widehat{M}(x,t)=i\hbar \partial _{t}+\mu -\frac{\widehat{p}^{2}}{2m}%
-U(x)-\phi (x,t)$ is the electron propagator operator in the new Hilbert
space, and $\mu $ is the chemical potential. The Lagrange multiplier field $%
\phi (x,t)$ is introduced to decouple the electron interaction, in the
meanwhile the original Hilbert space is mapped into a new Hilbert space, in
which the electrons are non-interacting, and moving in a fluctuating
potential produced by $\phi (x,t)$.

In general, the eigenequation of the electron propagator operator reads\cite%
{10,7},%
\begin{equation}
\widehat{M}(x,t)\Psi _{\sigma k\omega }(x,t;[\phi ])=E_{\sigma k\omega
}[\phi ]\Psi _{\sigma k\omega }(x,t;[\phi ])  \label{3}
\end{equation}%
and the eigenvalue can be obtained by the Hellmann-Feynman theorem,%
\begin{eqnarray*}
E_{\sigma k\omega }[\phi ] &=&\hbar \omega -E_{k}-\Sigma _{\sigma k}[\phi ]
\\
\Sigma _{\sigma k}[\phi ] &=&\int_{0}^{1}d\lambda \int dtd^{3}x\phi
(x,t)|\Psi _{\sigma k\omega }(x,t;[\lambda \phi ])|^{2}
\end{eqnarray*}%
where $\omega $ is the frequency, $\Sigma _{\sigma k}[\phi ]$ is the
self-energy of the electrons in the new Hilbert space, and $E_{k}$ is the
eigenvalue of the non-interacting Hamiltonian of the electrons,%
\begin{equation*}
H_{0}\psi _{\sigma k}(x)=(E_{k}+\mu )\psi _{\sigma k}(x)
\end{equation*}%
where $H_{0}=\frac{\widehat{p}^{2}}{2m}+U(x)$, and $k$ labels a set of
quantum number representing the states of the non-interacting electrons.
According to the expression of the eigenvalue $E_{\sigma k\omega }[\phi ]$,
the eigenfunctional can be generally written down,%
\begin{equation}
\Psi _{\sigma k\omega }(x,t;[\phi ])=\frac{1}{\sqrt{T}}\psi _{\sigma
k}(x)e^{-i(\omega -\Sigma _{\sigma k}[\phi ])t}e^{Q_{\sigma k}(x,t;[\phi ])}
\label{4}
\end{equation}%
where $T\rightarrow \infty $ is the time scale of the system, and the phase
field $Q_{\sigma k}(x,t;[\phi ])$ satisfies the eikonal-like equation with
the condition $Q_{\sigma k}(x,t;[\phi =0])=0$,%
\begin{equation}
\left( i\hbar \partial _{t}-\frac{\widehat{p}^{2}}{2m}-\frac{1}{m}[\widehat{p%
}\ln (\psi _{\sigma k}(x))]\cdot \widehat{p}\right) Q_{\sigma k}(x,t;[\phi
])-\frac{[\widehat{p}Q_{\sigma k}(x,t;[\phi ])]^{2}}{2m}=\phi (x,t)
\label{5}
\end{equation}%
For the homogeneous case, this equation can be easily solved after
neglecting the non-linear term which in general is a small quantity\cite{x}.
It is worthily noted that the difference between the eigenfunctional $\Psi
_{\sigma k\omega }(x,t;[\phi ])$ of the electrons in the new Hilbert space
and the wave function $\psi _{\sigma k}(x)$ of the non-interacting electrons
is the functional $e^{Q_{\sigma k}(x,t;[\phi ])}$ but a pure phase factor $%
e^{-i(\omega -\Sigma _{\sigma k}[\phi ])t}$, thus the physical meaning of
the phase field $Q_{\sigma k}(x,t;[\phi ])$ is very clear that the electron
correlation effect can be completely represented by the phase field, i.e.,
the phase field is a correlation parameter of the electrons. The quantum
phase transition of the system from the Landau Fermi liquid to the non-Fermi
(Luttinger) liquid can be described by the phase field.

In terms of the eigenfunctionals $\Psi _{\sigma k\omega }(x,t;[\phi ])$, the
second quantization representation of the electrons in the new Hilbert space
can be written down,%
\begin{eqnarray}
\widehat{\psi }_{\sigma }(x,t) &=&\sum_{k\omega }\Psi _{\sigma k\omega
}(x,t;[\phi ])\widehat{c}_{\sigma k\omega }  \notag \\
\widehat{\psi }_{\sigma }^{\dagger }(x,t) &=&\sum_{k\omega }\Psi _{\sigma
k\omega }^{\ast }(x,t;[\phi ])\widehat{c}_{\sigma k\omega }^{\dagger }
\label{6}
\end{eqnarray}%
where $\widehat{c}_{\sigma k\omega }$ ($\widehat{c}_{\sigma k\omega
}^{\dagger }$) is the annihilation (creation) operator of the electrons with
the spin index and the quantum number $k$ and $\omega $. Using the
orthogonality and the completeness of the eigenfunctionals $\Psi _{\sigma
k\omega }(x,t;[\phi ])$, it can be easily proved that the above electron
operators satisfy the standard electron anti-commutation relations. These
expressions play central role in calculating the ground state energy and a
variety of correlation functions of the system.

After integrating out the electron fields, and using the mathematical
formula,%
\begin{equation*}
\mathit{Tr\ln }\left( \widehat{A}+\widehat{B}\right) =\mathit{Tr\ln }\left( 
\widehat{A}\right) +\mathit{Tr}\int_{0}^{1}d\lambda \widehat{B}\frac{1}{%
\widehat{A}+\lambda \widehat{B}}
\end{equation*}%
the partition function can be written down\cite{9},%
\begin{eqnarray}
Z &=&\int D\rho D\phi e^{\frac{i}{\hbar }S[\rho ,\phi ]}  \notag \\
S[\rho ,\phi ] &=&-i\mathit{Tr\ln }\left( \widehat{M}_{0}\right) +W[\rho
,\phi ]+i\sum_{\sigma }\int_{0}^{1}d\lambda \int dtd^{3}x\phi (x,t)G_{\sigma
}(x,t;x,t;[\lambda \phi ])  \label{7}
\end{eqnarray}%
where $\widehat{M}_{0}=i\hbar \partial _{t}+\mu -\frac{\widehat{p}^{2}}{2m}%
-U(x)$ is the non-interacting electron propagator operator. The electron
Green function $G_{\sigma }(x,t;x^{\prime },t^{\prime };[\phi ])$ in the new
Hilbert space can be written down,%
\begin{equation*}
G_{\sigma }(x,t;x^{\prime },t^{\prime };[\phi ])=\sum_{k\omega }\frac{\Psi
_{\sigma k\omega }(x,t;[\phi ])\Psi _{\sigma k\omega }^{\ast }(x^{\prime
},t^{\prime };[\phi ])}{\hbar \omega -E_{k}-\Sigma _{\sigma k}[\phi ]}
\end{equation*}%
and the electron Green function in the original Hilbert space is that, $%
G_{\sigma }(x,t;x^{\prime },t^{\prime })=<G_{\sigma }(x,t;x^{\prime
},t^{\prime };[\phi ])>_{\rho \phi }$, where $<A[\phi ]>_{\rho \phi }=\frac{1%
}{Z}\int D\rho D\phi e^{\frac{i}{\hbar }S[\rho ,\phi ]}A[\phi ]$. The action 
$S[\rho ,\phi ]$ is used to calculate the functional average over the
auxiliary fields $\phi (x,t)$ and $\rho (x,t)$.

According to the expressions of the electron operators, the ground state
energy $E_{g}=<H(t)>$ of the system reads,%
\begin{equation}
E_{g}=T_{0}+V_{\text{HF}}+\Delta T+E_{c}  \label{8}
\end{equation}%
where $T_{0}=2\sum_{k}n_{k}\left( E_{k}+\mu \right) $ is the kinetic energy
of the non-interacting electrons, and $n_{k}=\theta (-E_{k})$ is the
occupation number of the $k$-th state of the non-interacting electrons. The
term $V_{\text{HF}}$ is the usual Hartree-Fock potential energy\cite{11},%
\begin{eqnarray}
V_{\text{HF}} &=&\frac{e^{2}}{2}\int d^{3}xd^{3}x^{\prime }V(x-x^{\prime
})\rho _{0}(x)\rho _{0}(x^{\prime })  \notag \\
&&-\frac{e^{2}}{2}\sum_{\sigma kk^{\prime }}n_{k}n_{k^{\prime }}\int
d^{3}xd^{3}x^{\prime }V(x-x^{\prime })\psi _{\sigma k}^{\ast }(x)\psi
_{\sigma k^{\prime }}(x)\psi _{\sigma k^{\prime }}^{\ast }(x^{\prime })\psi
_{\sigma k}(x^{\prime })  \label{9}
\end{eqnarray}%
where $\rho _{0}(x)=\sum_{\sigma k}n_{k}|\psi _{\sigma k}(x)|^{2}$ is the
density of the non-interacting electrons. The first two terms in the
expression of the ground state energy is similar to that in the LDA of the
density functional theory, which can be rigorously represented by the
single-particle wave functions.

The term $\Delta T$ is a modification of the kinetic energy of the electrons
by the Coulomb interaction\cite{12},%
\begin{equation}
\Delta T=\sum_{\sigma k}\int d^{3}x\rho _{\sigma k}(x)g_{\sigma k}(x,[\rho
_{0}])  \label{10}
\end{equation}%
where $\rho _{\sigma k}(x)=n_{k}|\psi _{\sigma k}(x)|^{2}$, and $g_{\sigma
k}(x,[\rho _{0}])=<|e^{Q_{\sigma k}(x,t;[\phi ])}|^{2}[i\hbar \partial
_{t}Q_{\sigma k}(x,t;[\phi ])-\phi (x,t)]>_{\rho \phi }$is the functional of
the density $\rho _{0}(x)$, and independent of the time coordinate $t$
because of the time translation symmetry of the system. For a homogeneous
electron system, the function $g_{\sigma k}(x,[\rho _{0}])=g_{\sigma k}[\rho
_{0}]$ is equivalent to the self-energy under the random-phase
approximation, and it is a functional of the density $\rho _{0}(x)$. Thus,
the term $\Delta T$ can be treated approximately under the LDA.

The correlation energy $E_{c}=E_{Hc}+E_{Fc}$ can be written down\cite%
{13,14,15},%
\begin{eqnarray}
E_{Hc} &=&\frac{e^{2}}{2}\int d^{3}xd^{3}x^{\prime }V(x-x^{\prime })\rho
(x,x^{\prime })  \notag \\
E_{Fc} &=&-\frac{e^{2}}{2}\int d^{3}xd^{3}x^{\prime }V(x-x^{\prime })\Gamma
(x,x^{\prime })  \label{11}
\end{eqnarray}%
where the functions $\rho (x,x^{\prime })$ and $\Gamma (x,x^{\prime })$ are
defined as, 
\begin{eqnarray*}
\rho (x,x^{\prime }) &=&\sum_{\sigma \beta kk^{\prime }}n_{k}n_{k^{\prime
}}|\psi _{\sigma k}(x)|^{2}|\psi _{\beta k}(x^{\prime })|^{2} \\
\cdot &<&|e^{Q_{\sigma k}(x,t;[\phi ])}|^{2}|e^{Q_{\beta k}(x^{\prime
},t;[\phi ])}|^{2}-1>_{\rho \phi } \\
\Gamma (x,x^{\prime }) &=&\sum_{\sigma kk^{\prime }}n_{k}n_{k^{\prime }}\psi
_{\sigma k}^{\ast }(x)\psi _{\sigma k^{\prime }}(x)\psi _{\sigma k^{\prime
}}^{\ast }(x^{\prime })\psi _{\sigma k}(x^{\prime }) \\
\cdot &<&e^{Q_{\sigma k}^{\ast }(x,t;[\phi ])+Q_{\sigma k}(x^{\prime
},t;[\phi ])+Q_{\sigma k^{\prime }}^{\ast }(x^{\prime },t;[\phi ])+Q_{\sigma
k^{\prime }}^{\ast }(x,t;[\phi ])}-1>_{\rho \phi }
\end{eqnarray*}%
Obviously, $E_{Hc}$ is the Hartree-like correlation energy, while $E_{Fc}$
represents the Fock-like correlation energy. The former shows the local
behavior, and the latter has the non-local property. It would be
demonstrated that the main errors of the LDA of the density functional
theory originates from the correlation energy $E_{Fc}$ term.

For a homogeneous interacting electron system, according to the experience
in studying the one-dimensional interacting electron systems\cite{xx}, the
electron correlation strength is controlled by the imaginary part of the
phase field $Q_{\sigma k}(x,t;[\phi ])$, and it can be shown that $%
<|e^{Q_{\sigma k}(x,t;[\phi ])}|^{2}|e^{Q_{\beta k}(x^{\prime },t;[\phi
])}|^{2}>_{\rho \phi }=e^{F_{\sigma \beta kk^{\prime }}(x-x^{\prime };[\rho
_{0}])}$, where the function $F_{\sigma \beta kk^{\prime }}(x-x^{\prime
};[\rho _{0}])$ is a small quantity, thus it is reasonable to take the
approximation\cite{xxx} $e^{F}\simeq 1+F$. Therefore, the function $\rho
(x,x^{\prime })$ can be approximately represented by the density function,%
\begin{equation}
\rho (x,x^{\prime })\simeq \sum_{\sigma \beta kk^{\prime }}F_{\sigma \beta
kk^{\prime }}(x-x^{\prime };[\rho _{0}])\rho _{\sigma k}(x)\rho _{\beta
k^{\prime }}(x^{\prime })  \label{12}
\end{equation}%
where $F_{\sigma \beta kk^{\prime }}(x-x^{\prime };[\rho _{0}])$ is the
functional of the non-interacting electron density $\rho _{0}(x)$. Thus, the
correlation energy $E_{Hc}$ can be treated approximately under the LDA.

It can be shown that $<e^{Q_{\sigma k}^{\ast }(x,t;[\phi ])+Q_{\sigma
k}(x^{\prime },t;[\phi ])+Q_{\sigma k^{\prime }}^{\ast }(x^{\prime },t;[\phi
])+Q_{\sigma k^{\prime }}^{\ast }(x,t;[\phi ])}>_{\rho \phi }=e^{P_{\sigma
kk^{\prime }}(x-x^{\prime };[\rho _{0}])}$ for the homogeneous case, while
the function $P_{\sigma kk^{\prime }}(x-x^{\prime };[\rho _{0}])$ is not a
small quantity\ for the systems with the strong electron correlation. For
example, for one-dimensional interacting electron gas, the function $%
e^{P_{\sigma kk^{\prime }}(x-x^{\prime };[\rho _{0}])}$ has a power-law
asymptotic behavior\cite{16},%
\begin{equation}
e^{P_{\sigma kk^{\prime }}^{\text{1D}}(x-x^{\prime };[\rho _{0}])}\sim \frac{%
1}{|x-x^{\prime }|^{\gamma }}  \label{13}
\end{equation}%
where $\gamma >1$ is a dimensionless coupling constant which depends upon
the electron interaction strength and the Fermi velocity. In this case, the
function $\Gamma (x,x^{\prime })$ shows completely different behavior from
that of the function $\rho (x,x^{\prime })$, and the correlation energy $%
E_{Fc}$ cannot be treated approximately under the LDA. In general, it is
expected that the correlation energy $E_{Fc}$ is important for the systems
with the strong electron correlation, and it dominates the correlation
energy of the systems, thus a method beyond the LDA is needed.

The above expression of the ground state energy is obviously different from
that of the density functional theory in the LDA (Kohn-Sham scheme) where
the ground state energy is represented by the quantity of the
non-interacting electrons. The ground state energy in equation (\ref{8}) is
composed of two parts, one part is corresponding to the non-interacting
electrons, and another one is represented by the correlation functions that
are controlled by the electron correlation which describes the many-electron
effect. In principle, the expression of the ground state energy is
independent of the choice of the Hamiltonian $\widehat{H}_{0}$ of the
non-interacting electrons. The different choice of the Hamiltonian $\widehat{%
H}_{0}$ can give different non-interacting part, while for the correlation
part, it will only change the action $S[\rho ,\phi ]$ and the wave function $%
\psi _{\sigma k}(x)$ of the non-interacting electrons.

Based upon the above discussions, a more effective method for calculating
the ground state energy $E_{g}$ is that, by suitably choosing the
non-interacting Hamiltonian $\widehat{H}_{0}$, the terms $T_{0}$, $V_{\text{%
HF}}$, $\Delta T$ and $E_{Hc}$ can be calculated with usual LDA scheme, then
the term $E_{Fc}$ can be calculated by the quantum Monte Carlo method (QMCM)%
\cite{17} where the action $S[\rho ,\phi ]$ can be obtained by solving the
equation (\ref{5}) of the phase field. By more accurately treating the
correlation energy $E_{Fc}$, the results of usual LDA will be heavily
modified for the systems with strong electron correlation.

On the other hand, for a suitable LDA, if the contribution of the
correlation energy $E_{Fc}$ to the ground state energy is small, this LDA is
reliable. If its contribution is important, a new LDA has to be chosen to
make its contribution be as small as possible. Of course, corresponding to
the local spin density approximation (LSDA)\cite{15} of the density
functional theory, it needs to introduce the Lagrange multiplier fields $%
\phi _{\sigma }(x,t)$, in which way there are four auxiliary fields $\phi
_{\sigma }(x,t)$ and $\rho _{\sigma }(x,t)$, while the ground state energy
has the similar expression as that in equation (\ref{8}).

In summary, with the eigenfunctional theory, we have studied a general
interacting electron system, and given a rigorous expression of its ground
state energy which is composed of two parts, one part is contributed by the
non-interacting electrons, and another one is represented by the correlation
functions that are controlled by the electron correlation. Moreover,
according to the rigorous expression of the ground state energy, an
effective method beyond LDA may be the LDA plus the QMCM. The LDA is used to
treat the non-interacting electron part, while the correlation part is
treated by the QMCM.

This work is supported by the NSF of China, No.90403015 and No.10774188.


\begin{thebibliography}{99}
\bibitem{x} It can be easily understood for the lattice case where the
non-linear term originates from the phase difference between the nearest
neighbour sites. In general, the phase field produced by the quantum
fluctuation of the Lagrange multiplier field is a smooth function, thus the
non-linear term is a small quantity.
\bibitem{xx} In usual bosonization method, the phase field is a pure
imaginary function.
\bibitem{xxx} According to the normalization of the eigenfunctional, there
is the relation, $<|e^{Q_{\sigma k}(x,t;[\phi ])}|^{2}>_{\rho \phi }=1$.
\end{thebibliography}
\end{document}